\documentclass[12pt,preprint]{aastex}

\slugcomment{To appear in {\em ApJ Lett.}}
\begin{document}
%\title{A beaming factor of 450 in GRBs with standard energies}
\title{Observational evidence for a correlation between 
peak-luminosities and beaming in GRBs}
\author{Maurice H.P.M. van Putten\altaffilmark{1} and Tania Regimbau\altaffilmark{1}}
\affil{LIGO Laboratory, NW 17-161, 175 Albany Street, Cambridge, MA 02139}
\begin{abstract}
We calculate the unseen-but-true GRB-event rate from a current flux-limited sample of 33 
GRBs with individually measured redshifts. We consider a GRB-event rate which is proportional
to the star-formation rate, in view of the GRB-SNe association in SN1998bw and GRB030329.
By fitting a log-normal distribution of GRB peak-luminosities, 
we find a ratio $1/f_r\simeq450$ 
for the true-to-observed GRB-event rates. This provides an independent derivation 
of the GRB-beaming factor $1/f_b\simeq500$ obtained by Frail et al. (2001) from 
sources with standard GRB-energies. We discuss applications to GRB980425.
\end{abstract}
\section{Introduction}
New clues are emerging that long GRB are associated with supernovae 
\citep{gal98,blo99,kul00,rei01}. This is illustrated most recently 
by optical emissions lines in the late-time light-curve of the HETE-II burst 
GRB 030329 \citep{sta03}, which are remarkably similar to those observed
in GRB980425/SN1998bw \citep{gal98}. A GRB-supernova assocation provides important
support for the collapsar model of GRBs, representing a violent death
of evolved massive stars \citep{woo93,pac98}. The short lifespan of tens of Myrs
of massive stars implies that GRBs take place in star-forming regions
\citep{pac98,fru99}, and hence more broadly points towards an association
to molecular clouds. The event rate of GRBs per unit cosmological volume 
is hereby expected to be correlated to the cosmic star-formation rate
(e.g., \citet{bla00,ber03,cho02}). Because 
observations by past and current experiments (see Table I) 
are flux-limited, the observed GRB-event rate is strongly biased towards 
events at lower redshifts.

In this letter, we calculate the ratio $1/f_r$ of unseen-to-observed GRBs.
Our analysis is based on the bias in the observed GRB-event rate towards 
low redshift events in a flux-limited sample of 33 GRBs with individually
measured redshifts. We propose to use a linear relationship for the correlation
between the true-but-unseen GRB event rate and the cosmic SFR, to derive from
this bias a best-fit log-normal GRB peak-luminosity function. While it is an 
open question to what the degree cosmological evolution of metallicity 
affects the SFR independently of the GRB event rate (see, e.g., \cite{heg02}),
linear relationship between the two will serve as a leading-order approximation.

In \S2, we discuss the cosmological model. \S3 tabulates the current sample
of 33 GRBs with individually measured redshifts. In \S4, we present the
best fit of the data for a log-normal distribution function, estimate
the fraction of unseen-to-observed GRBs, and compare our results with
the beaming factor $1/f_b=500$ of Frail et al.(2001), 
based on a sample of GRBs with achromatic breaks
in their light-curves. We summarize our results and draw our main conclusions 
in \S5. 

\section{The cosmological model}

We model the intrinsic GRB-event rate in a flat $\Lambda$-dominated cold
dark matter cosmology with closure energy densities $\Omega_\Lambda$ = 0.70
and $\Omega_m$ = 0.30. These values are suggested by BOOMERANG and MAXIMA 
\citep{deb00,han00} on the power spectra of the CMB and distant Type Ia SNe
\citep{per99,sch98}. The Hubble parameter H$_o$ is taken to be 
$73\,\mathrm{km\,s}^{-1}\mathrm{Mpc}^{-1}$ (Freedman et al., 2001).

\citet{por01} provide three models of the cosmic SFR up to redshifts $z\sim5$,
reflecting some uncertainties in SFR estimates.
No significant changes have been noticed in application of these three models,
and we present only results derived from the second model, SFR2. We have 
\begin{equation}
R_{c}(z) = 
R_{SF2}(z)
\frac{E(\Omega_i,z)}{(1+z)^{3/2}}\; 
\mathrm{M}_{\odot}\,\mathrm{yr}^{-1}\,\mathrm{Mpc}^{-3}
\end{equation}
by transformation of the $R_{SF2}$ in a matter-dominated
unverse ($\Omega_{m}=1$), 
\begin{equation}
R_{SF2}(z) = \frac{0.15\,e^{3.4z}}{(22 + e^{3.4z})} \; \mathrm{M}_{\odot}\,\mathrm{yr}^{-1}\,\mathrm{Mpc}^{-3},
\end{equation}
where $E(\Omega_i,z) =[(1+z)^2(1+z\Omega_m)-z(2+z)\Omega_v]^{1/2}$.

For a GRB-event rate locked to the SFR, 
the true event rate between $z$ and $z+dz$, as observed in the observer's frame 
of reference, satisfies
\begin{equation}
dR_{GRB}(z)=\lambda _{GRB}\frac{R_{c}(z)}{1+z}{\frac{{dV}}{{dz}}}\,dz
\end{equation}
where $\lambda_{GRB}$ is the formation mass fraction of the source
progenitors. Here, the division by $1+z$ 
accounts for time-dilatation by cosmic expansion.
The element of comoving volume is
\begin{equation}
dV = \frac{4\pi r^{2}c}{H_0E(\Omega_i,z)}\,dz,~~~
r (z) = \int_{0}^{z} \frac{c}{H_0E(\Omega_i,z')}\,dz'
\end{equation}
The GRB redshift probability density can be written as 
\citep{cow02}
%\begin{equation}
$p (z) =  {dR_{GRB}/dz}\left({\int_{0}^{5}dR_{GRB}/dz\,dz}\right)^{-1}.$
%\end{equation}
Assuming that the mass fraction of GRBs progenitors is redshift independent, 
the scaling factor $\lambda _{GRB}$ is the only free parameter of our model.
For the flux-limited experiments listed in Table I
we define a probability-density function of detection as a function of
redshift as
%\begin{equation}
$p_{detect} (z) ={dR_{detect}/dz}\left({\int_{0}^{5}dR_{detect}/dz\,dz}\right)^{-1},$
%\end{equation}
where the dependence on the luminosity has been integrated out in the
detected event rate \citep{bro02}
\begin {equation}
dR_{detect} = dR_{GRB}(z) \int_{L_{lim}(z)} p(L)\,dL
\end {equation}
Here, $p(L)$ refers to the intrinsic GRBs luminosity function
in the BATSE energy range $50 - 300$ keV. 

The luminosity threshold as a function of redshift is given by
%\begin {equation}
$L_{lim} (z) = 4\pi d_{L}^{2}(z) S_{lim}$,
%\end {equation}
where $d_{L}$ is  the luminosity distance to a source at redshift $z$ and
where $S_{lim}$ denotes the sensitivity threshold of the
instrument.  Following \citep{bro02}, we take a flux-density
threshold of BASTE of 0.2 $\mathrm{photon}\,\mathrm{cm}^{-2}\,\mathrm{s}^{-1}$.

\section{A redshift sample of 33 GRBs}

As of the time of this writing, we have 33 GRBs with
individually measured redshifts (Table I). They collectively
represent a variety of past and current experiments.
Due to a flux threshold in each of the instruments used, 
the observed GRB redshift distribution is strongly biased
towards low redshifts. This introduces a quantitative
selection effect, relative to the redshift distribution 
predicted by the SFR model (Fig. 1) -- the true distribution 
of GRB-redshifts representing what would be observed in the 
ideal case of a zero-flux threshold in the instruments.
In view of the various instruments involved, the dependency of
the observed GRB redhift-distribution function on the flux-limit
is simulated in Fig. 2.

%\section{The GRB-luminosity function}
%We set out to derive the intrinsic GRBs luminosity function $p(L)$
%by comparing the true GRB-redshift distribution locked to the SFR 
%and the observed GRB-redshift distribution, listed in Table I.

\section{A fit to a log-normal distribution function}

We shall assume that the GRB-luminosity function is redshift independent,
i.e., without cosmological evolution of the nature of its progenitors.
We take a log-normal probability density for the luminosity shape-function,
with mean $\mu$ and width $\sigma$ given by
\begin{equation}
p(L) = \frac{1}{(2\pi)^{1/2} \sigma L}\exp
\left(\frac{-(\ln L-\mu)^{2}}{2\sigma^{2}}\right),
\label{EQN_LNP}
\end{equation}
where $L$ is normalized with respect to  1cm$^{-2}$s$^{-1}$.
The optimized parameters of our model are
\begin{eqnarray}
(\mu , \sigma) = (124 , 3)
\pm (2 , -0.4).
\label{EQN_PAR}
\end{eqnarray}
This notation means that the estimated parameters can be either 
(122,3.4), (123,3.2), (125,2.8), or (126,2.6), but not (122,2.6)
for instance. Our results are compatible to the expectations
of Sethi et Bhargavi (2001) who derive a log-normal luminosity
function with $\mu = 129$ and $\sigma = 2$ from a different flux-limited
sample. They mentioned that because of selection effects, the
inferred average luminosity over-estimates the true mean luminosity 
by a factor of 2 or 3, and the variance is within 40-50\% of the true
variance. The observed and predicted redshift distribution, based on the SFR,
are shown in Fig. 1 in case of optimal parameters (\ref{EQN_PAR}).
The results indicate a good fit to the data, suggesting that selection effects 
are adequaly modeled. The fraction of 
{\em detectable} GRBs as a function of redshift
\begin{equation}
F(z) = \frac{dR_{detect}}{dR_{GRB}(z)} =  \int_{L_{lim}(z)} p(L)\,dl .
\end{equation}
shows a the steep decrease as the luminosity threshold
increases, making high-redshift GRBs less likely to be detected.  

A satisfactory test for our model is provided by 
comparing the predicted flux distribution
of GRBs above the sensitivity threshold with the peak flux distribution of
a sample of 67 GRBs observed with IPN (Fig. 3).
The fluxes derived from our luminosity function in 50-300 keV have been 
extrapolated to the IPN range of 25-100 keV, assuming an $\mathrm{E}^{-2}$ 
energy spectrum and using the formula given in Appendix B of
Sethi et Bhargavi (2001). 
The conversion factor from $\mathrm{erg}\,\mathrm{cm}^{-2}\,\mathrm{s}^{-1}$ to
$\mathrm{photon}\,\mathrm{cm}^{-2}\,\mathrm{s}^{-1}$ has been taken
to be $0.87 \times 10^{-7}$, and the sensitivity
threshold equal to 
5 $\mathrm{photon}\,\mathrm{cm}^{-2}\,\mathrm{s}^{-1}$ (Hurley, private
communication).

A quantitative result provided by our model is the ratio $1/f_r$
of the true GRB-event rate (what would be seen in case of a zero-flux limit
in the detector) and the observed GRB-event rate. For the optimal
parameters (\ref{EQN_PAR}), we have
\begin{eqnarray}
1/f_r = 450.
\label{EQN_R}
\end{eqnarray}
The factor $1/f_r$ is between 200-1200 in the error box of (\ref{EQN_PAR}).

\section{A correlation between $1/f_r$ and $1/f_b$}

Our estimate of the true-to-observed GRB-event rate $1/f_r$ is strikingly 
similar to the GRB-beaming factor $1/f_b$ of about 500 derived by Frail et al. 
(2001).
Our analysis is independent of the mechanism providing a broad distribution 
in GRB luminosities. Without further input, our results may reflect 
(a) isotropic sources with greatly varying energy output, 
(b) beamed sources with standard energy output and varying opening angles, or 
(c) anisotropic but geometrically standard sources \citep{ros02,zha02}.

We find that the GRB peak-luminosities and beaming are correlated. 
To see this, we simply note that the case of no correlation between 
peak-luminosities and beaming give rise to an unseen-but-true GRB event rate 
which is $1/f_r\times 1/f_b\simeq 2.5\times10^5$ times the observed rate. The 
true GRB-event rate hereby approaches that of Type II supernovae -- we discard 
this possibility.  A correlation between peak-luminosities and beaming is 
naturally expected sources (b) and (c) with standard energy output -- the 
picture that bears out of \citet{fra01}. This introduces a correlation 
$L\propto 1/f_b$ between peak-luminosity and beaming factor in (b) and, for 
a flux-limited sample, also in (c). For a flux-limited sample, both (b) and
(c) give rise to an anticorrelation between inferred beaming and distance 
such that to leading-order $\theta_jz\sim$const. Fig. 4 shows that this 
anticorrelation holds in the sample of \citet{fra01}. 
For a related discussion on estimating the GRB beaming factor from
flux limited surveys, see \citet{lev02}.

\section{Conclusions}

A table of 33 GRBs with individually determined
redshifts allows an estimate of the GRB-luminosity function, based
on constant of proportionality between the GRB event rate and the 
cosmic star-formation rate. We have tested our fit by reproducing the 
distribution of peak luminosities in the IPN sample of 67 GRBs.

A flux-limited sample introduces a ratio $1/f_r$ of unseen GRBs,
whose emissions fall below the detector threshold, to observed GRBs. 
A best fit analysis of the luminosity function gives $1/f_r\simeq450$. 
This number is both large and close to the GRB beaming factor of 500. 
The possibility that $1/f_r$ and $1/f_b$ are uncorrelated 
gives rise to prohibitively large true GRB event rates, and is discarded.
The results support a relation $L\propto 1/f_b$ between the peak-luminosity 
and the beaming factor in the context of conical or anisotropic GRB-emissions 
with standard GRB-energies. This points towards an anticorrelation between 
distance and half-opening angle, which is approximately supported by the 
sample of \citet{fra01}.

The results presented here and those of \citet{fra01} indicate a true
GRB event rate of 1 per year within $D=100$Mpc. 
GRB980425/SN1998bw ($z=0.0085$) is consistent with this true
event rate. For anisotropic but geometrically standard sources, 
GRB980425 is hereby {\em not} anomalous, but 
consistent with the trend shown in Fig. 4.

\acknowledgments

The authors greatfully acknowledge input from Kevin Hurley. 
MVP thanks R. Mochkovitch for discussions on GRB980425.
This research is
supported by the LIGO Observatories, constructed by Caltech and MIT
with funding from NSF under cooperative agreement PHY 9210038.
The LIGO Laboratory operates under cooperative agreement
PHY-0107417. This paper has been assigned LIGO document number
LIGO-P030022-00-D.

\begin{deluxetable}{crrrrr}
\tabletypesize{\scriptsize}
\tablecaption{A sample of 33 GRBs with individually determined redshifts
\tablenotemark{a}\label{tbl-1}}
\tablewidth{0pt}
\tablehead{
%\colhead{GRB} 
GRB & Redshift $z$ & Photon flux\tablenotemark{b} & 
Luminosity \tablenotemark{c}& $\theta_j$\tablenotemark{d} & Instrument}
\startdata
970228  & 0.695 & 10      & $2.13 \times 10^{58}$ &         & SAX/WFC\\

970508  & 0.835 & 0.97    & $3.24 \times 10^{57}$ & 0.293   & SAX/WFC\\

970828  & 0.9578& 1.5     & $7.04 \times 10^{57}$ & 0.072   & RXTE/ASM\\

971214  & 3.42  & 1.96    & $2.08 \times 10^{59}$ & $>0.056$& SAX/WFC\\

980425  & 0.0085& 0.96    & $1.54 \times 10^{53}$ &         & SAX/WFC\\

980613  & 1.096 & 0.5     & $3.28 \times 10^{57}$ & $>0.127$& SAX/WFC\\

980703  & 0.966 & 2.40    & $1.15 \times 10^{58}$ & 0.135   & RXTE/ASM\\

990123  & 1.6   & 16.41   & $2.74 \times 10^{59}$ & 0.050   & SAX/WFC\\

990506  & 1.3   & 18.56   & $1.85 \times 10^{59}$ &         & BAT/PCA\\

990510  & 1.619 & 8.16    & $1.40 \times 10^{59}$ & 0.053   & SAX/WFC\\

990705  & 0.86  &         &                     & 0.054   & SAX/WFC\\

990712  & 0.434 & 11.64   & $7.97 \times 10^{57}$ & $>0.411$& SAX/WFC\\

991208  & 0.706 & 11.2*   & $2.48 \times 10^{58}$ & $<0.079$& Uly/KO/NE\\

991216  & 1.02  & 67.5    & $3.70 \times 10^{59}$ & 0.051   & BAT/PCA\\

000131  & 4.5   & 1.5*    & $3.05 \times 10^{59}$ &$<0.047$ & Uly/KO/NE\\

000210  & 0.846 & 29.9    & $1.03 \times 10^{59}$ &         & SAX/WFC\\

000301C & 0.42  & 1.32*   & $8.37 \times 10^{56}$ & 0.105   & ASM/Uly\\

000214  & 2.03  &         &                     &         & SAX/WFC\\

000418  & 1.118 & 3.3*    & $2.27 \times 10^{58}$ & 0.198   & Uly/KO/NE \\

000911  & 1.058 & 2.86    & $1.72 \times 10^{58}$ &         & Uly/KO/NE\\

000926  & 2.066 & 10*     & $3.13 \times 10^{59}$ & 0.051   & Uly/KO/NE\\

010222  & 1.477 &         &                     &         & SAX/WFC\\

010921  & 0.45  &         &                     &         & HE/Uly/SAX\\

011121  & 0.36  & 15.04*  & $6.63 \times 10^{57}$ &         & SAX/WFC\\

011211  & 2.14  &         &                     &         & SAX/WFC\\

020405  & 0.69  & 7.52*   & $1.58 \times 10^{58}$ &         & Uly/MO/SAX\\

020813  & 1.25  & 9.02*   & $8.19 \times 10^{58}$ &         & HETE\\

021004  & 2.3   &         &                     &         & HETE\\

021211  & 1.01  &         &                     &         & HETE\\

030226  & 1.98  & 0.48*   & $1.35 \times 10^{58}$ &         & HETE\\

030323  & 3.37  & 0.0048* & $4.91 \times 10^{56}$ &         & HETE\\

030328  & 1.52  & 2.93*   & $4.31 \times 10^{58}$ &         & HETE\\

030329  & 0.168 & 0.0009* & $7.03 \times 10^{52}$ &         & HETE
\enddata
\tablenotetext{a}{Compiled from S. Barthelmy's IPN redshifts and fluxes
(http://gcn.gsfc.nasa.gov/gcn/) and J.C. Greiner's catalogue on GRBs
localized with WFC (BeppoSax), BATSE/RXTE or ASM/RXTE, IPN, HETE-II
or INTEGRAL (http://www.mpe.mpg.de/~jcg/grbgeb.html)}
\tablenotetext{b}{in cm$^{-2}$s$^{-1}$}
\tablenotetext{c}{Photon luminosities in s$^{-1}$ derived from the measured redshifts and observed gamma-ray fluxes for the cosmological model described in \S2}
\tablenotetext{d}{Opening angles $\theta_j$ in the GRB-emissions refer to the 
sample listed
in Table I of Frail et al.(2001).}
\tablenotetext{*}{Extrapolated to the BATSE energy range 50 - 300 keV using the formula given in 
Appendix B of Sethi et Bhargavi (2001)}
\end{deluxetable}

\newpage
\centerline{Figure Captions}
%\mbox{}\\
%\mbox{}\\
%{\bf Figure 1.} Shown is the luminosity threshold as a function of redshift for the 
%BATSE flux-threshold of 0.2 $\mathrm{photon}\,\mathrm{cm}^{-2}\,\mathrm{s}^{-1}$,
%where $L_{\mbox{\tiny min}}$ is in units of cm$^{-2}$s$^{-1}$.
\mbox{}\\
\mbox{}\\
{\bf Figure 1.} Shown are three redshift distributions: the observed sample derived
from Table 1 (white), the true sample assuming the GRB event rate is 
locked to the star-formation rate (hachured), and the sample of detectable 
GRBs predicted by our model according to a log-normal peak-luminosity distribution
function (grey). The continuous line represents the cosmic star formation rate
according to a $\Lambda-$dominated cold dark matter universe.
\mbox{}\\
\mbox{}\\
{\bf Figure 2.} Shown is a simulation of the redshift distribution of the
observed GRBs as a function of flux-limit, corresponding to various
instruments including the upcoming SWIFT mission. The results are  
derived assuming the GRB event rate to be locked to the star-formation
rate, using the best fit log-normal peak-luminosity distribution function
used in Fig. 2. HETE-II tresholds are 0.21 (SXC), 0.07 (WXM) and 0.3 (FREGATE)
in units of cm$^{-2}$s$^{-1}$.
\mbox{}\\
\mbox{}\\
{\bf Figure 3.} Shown is a comparison between the flux distributions
derived from our model (in grey) and from a sample of 67 GRBs
observed with IPN (in wite). The fit serves as a test for
our model assumptions, namely a log-normal GRB-luminosity 
function and a GRB event rate locked to the star-formation rate.
\mbox{}\\
\mbox{}\\
%{\bf Figure 5.} Shown is the fraction of detectable GRBs as a function of redshift 
%for the optimal parameters in fitting a log-normal GRB-luminosity function.
%\mbox{}\\
%\mbox{}\\
{\bf Figure 4.} Shown is a plot of the opening angle $\theta_j$ of GRB-emissions
versus redshift $z$ in the sample of \citet{fra01}, as derived from achromatic breaks 
in the GRB light curves. These results indicate an anticorrelation
between $\theta_j$ and $z$. For standard GRB-energies, this introduces a peak-luminosity
function of GRBs which is correlated with the beaming factor $1/f_b$. This 
allows the beaming factor to be determined also in terms of the ratio of the
unseen-but-true GRB event rate to the observed GRB event rate, using the current 
flux-limited sample of 33 GRBs with individually measured redshifts.

\newpage
%\begin{figure}
%\plotone{fig1}
%\end{figure}

\begin{figure}
\plotone{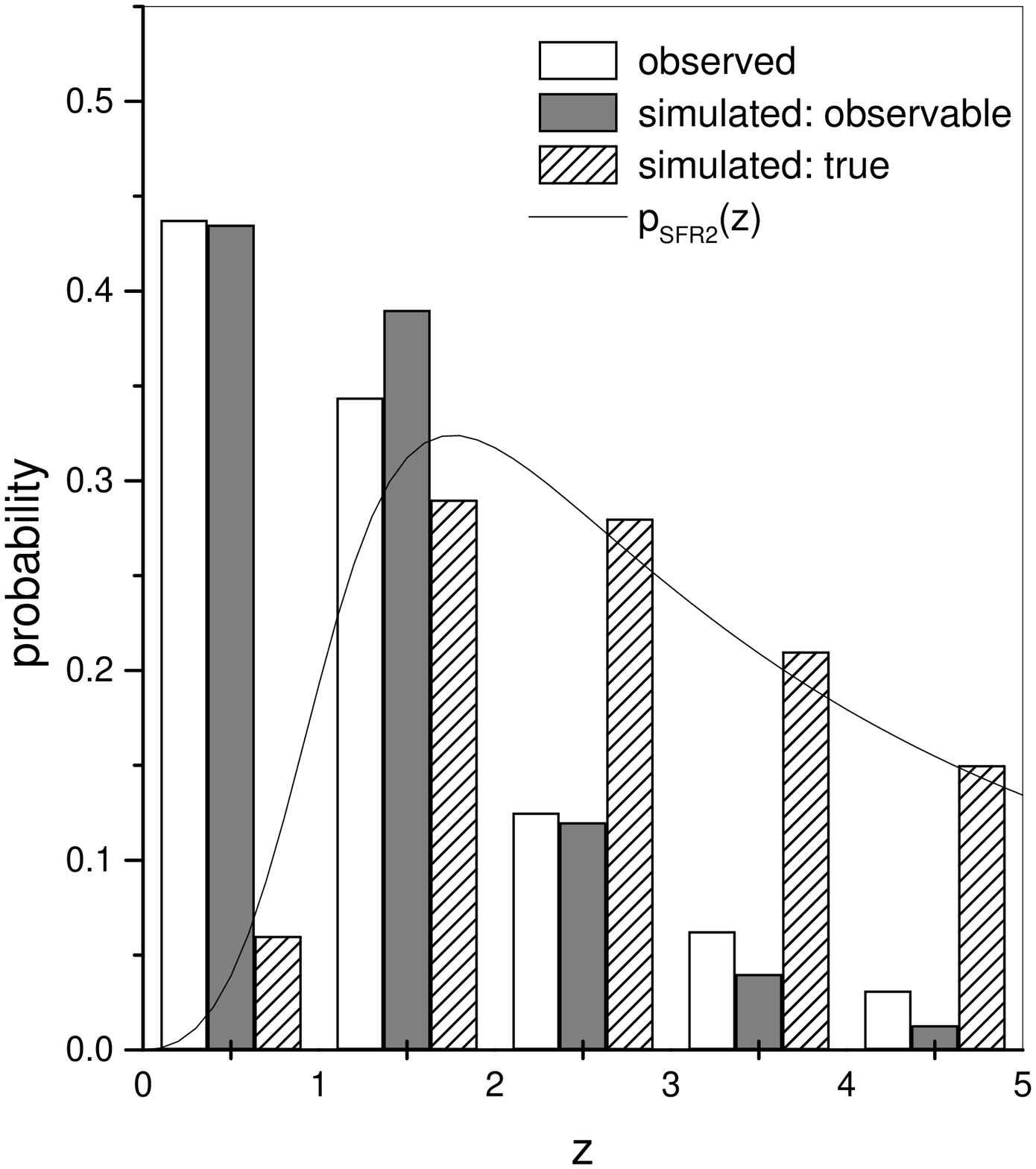}
\caption{}
\end{figure}

\begin{figure}
\plotone{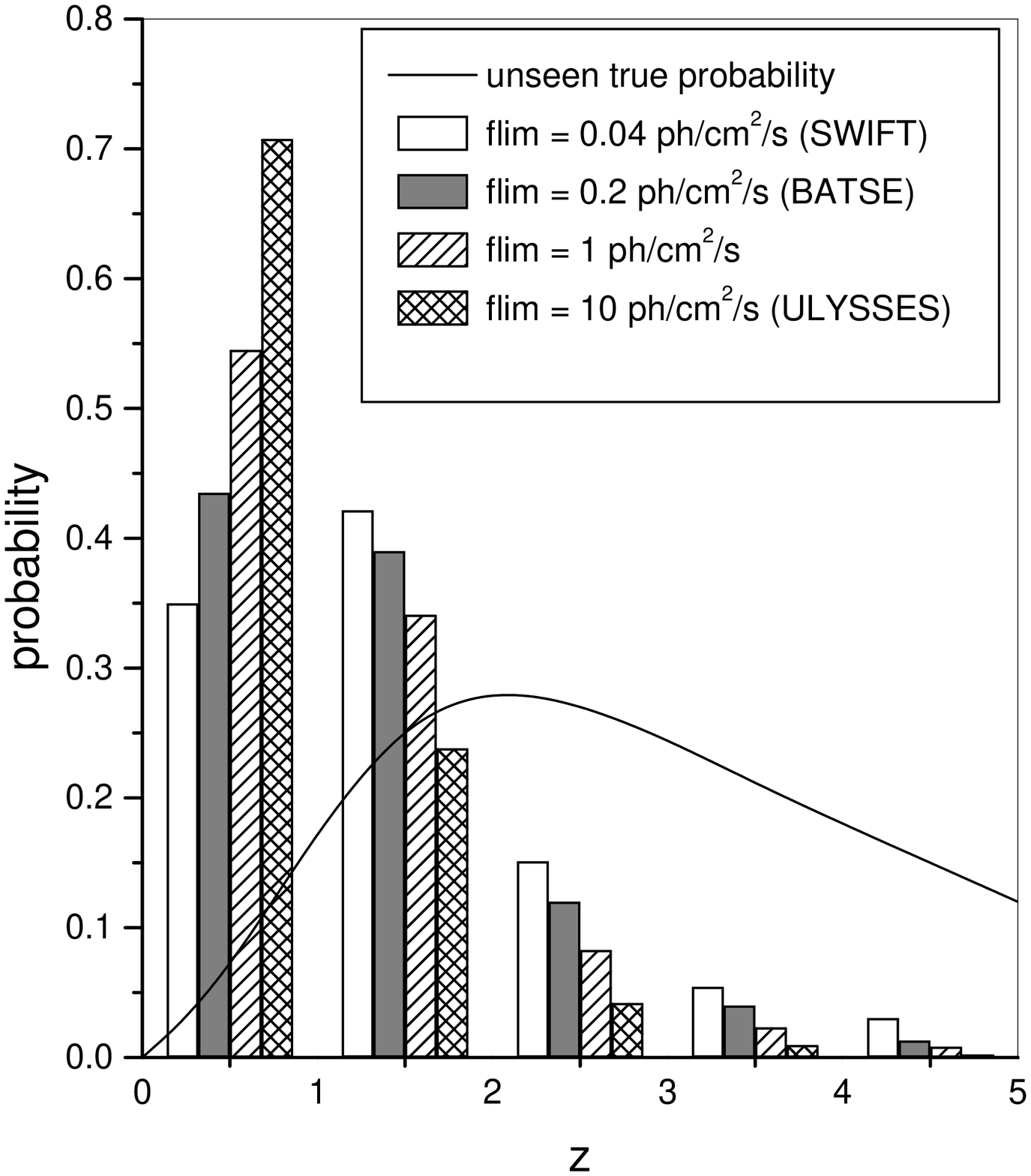}
\caption{}
\end{figure}

\begin{figure}
\plotone{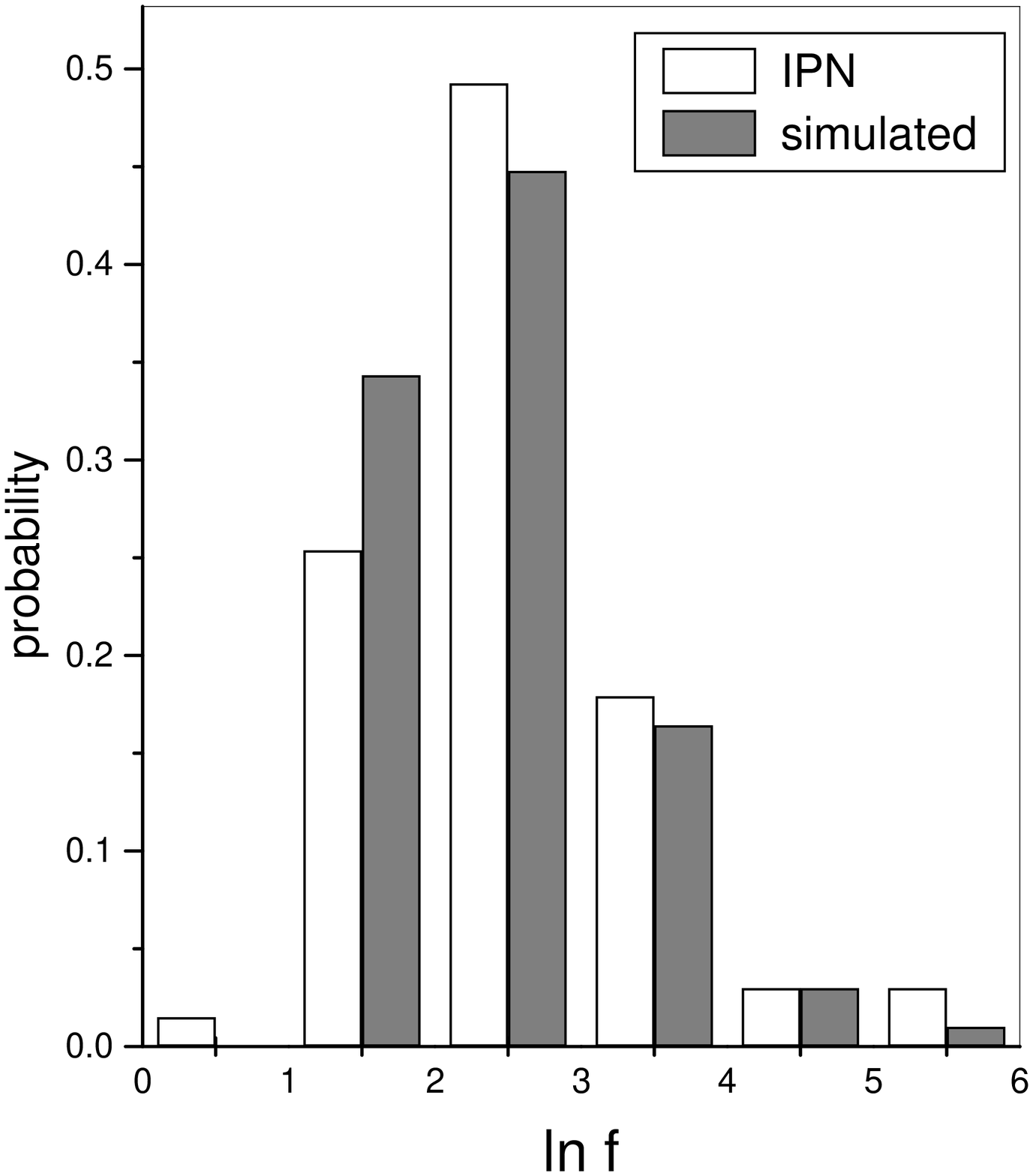}
\caption{}
\end{figure}

%\begin{figure}
%\plotone{fig5}
%\end{figure}

\begin{figure}
\plotone{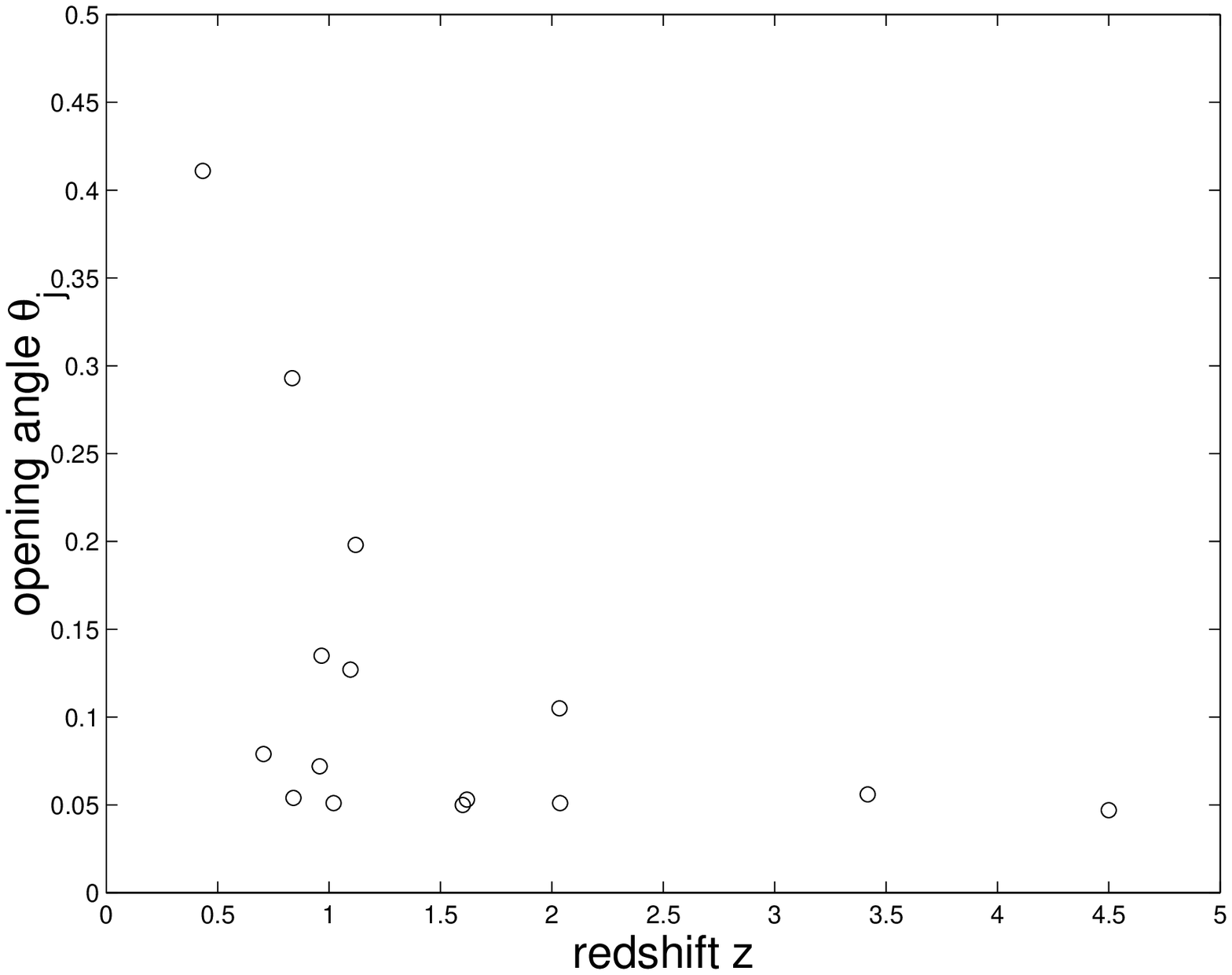}
\caption{}
\end{figure}

\end{document}